\documentclass[12pt,manuscript=article]{achemso}
\setkeys{acs}{maxauthors = 0}
\usepackage{graphics,epsfig,color,amsmath}
\usepackage{subcaption}
\usepackage{comment}
\usepackage{setspace}
\usepackage{siunitx}
\usepackage{hyperref}

\title{Binding Selectivity Analysis from Alchemical Receptor Hopping and Swapping Free Energy Calculations}

\author{Solmaz Azimi}
\affiliation{Department of Chemistry and Biochemistry, Brooklyn College of the City University of New York, New York, NY, 11210}
\alsoaffiliation{Ph.D. Program in Biochemistry, The Graduate Center of the City University of New York, New York, NY, 10016}
\author{Emilio Gallicchio}
\email{egallicchio@brooklyn.cuny.edu}
\affiliation{Department of Chemistry and Biochemistry, Brooklyn College of the City University of New York, New York, NY, 11210}
\alsoaffiliation{Ph.D. Program in Chemistry, The Graduate Center of the City University of New York, New York, NY, 10016}
\alsoaffiliation{Ph.D. Program in Biochemistry, The Graduate Center of the City University of New York, New York, NY, 10016}
\date{February 2024}

\begin{document}

\maketitle
\begin{abstract}
We present receptor hopping and receptor swapping free energy estimation protocols based on the Alchemical Transfer Method (ATM) to model the binding selectivity of a set of ligands to two arbitrary receptors. The receptor hopping protocol, where a ligand is alchemically transferred from one receptor to another in one simulation, directly yields the ligand's binding selectivity free energy for the two receptors, which is the difference between the two individual binding free energies. In the receptor swapping protocol, the first ligand of a pair is transferred from one receptor to another while the second ligand is simultaneously transferred in the opposite direction. The receptor swapping free energy yields the differences in binding selectivity free energies of a set of ligands, which, when combined using a generalized DiffNet algorithm, yield the binding selectivity free energies of the ligands. We test these algorithms on host-guest systems and show that they yield results that agree with experimental data and are consistent with differences in absolute and relative binding free energies obtained by conventional methods. Preliminary applications to the selectivity analysis of molecular fragments binding to the trypsin and thrombin serine protease confirm the potential of the receptor swapping technology in structure-based drug discovery. The novel methodologies presented in this work are a first step toward streamlined and computationally efficient protocols for ligand selectivity optimization across protein receptors with potentially low sequence identity.
\end{abstract}

\section{Introduction}

In drug discovery research, once an initial high-affinity inhibitor has been identified, further lead optimization typically includes evaluating the compounds' receptor selectivity profile to avoid unintended off-target activity or, conversely, broadening the activity spectrum to, for example, protect against resistance mutations.\cite{janes2018targeting,heffron2011pi3k,sutherlin2012pi3k} Achieving the desired selectivity profile often leads to a drug with fewer side effects, lower toxicity, and longer-lasting therapeutic potency. Inhibitor selectivity optimization is a very challenging process that requires input from multiple experimental assays to find the best balance between potency and specificity.\cite{schonherr2024discovery,heffron2016pi3k,estrada2012lrrk2} In this context, structure-based computational modeling can offer unique atomistic-level insights on the ligand modifications more likely to leverage energetic and structural differences between the target and homologous receptors.\cite{hauser2018predicting,moraca2019pde,fujimoto2021bace12,chen2012lrrk2}

 The selectivity of an inhibitor for a target receptor relative to an off-target receptor is measured quantitatively by the selectivity coefficient defined as the ratio of the binding constant of the inhibitor for the target receptor to its binding constant for the off-target receptor, with higher ratios indicating higher selectivity.\cite{klotz1997ligand,bayly1999structure} Because of the $\Delta G_b^\circ = -k_B T \ln K_b$ relationship between the binding constant, $K_b$, and the standard binding free energy, $\Delta G_b^\circ$, the difference between the standard binding free energies of the inhibitor to the target receptor relative to a reference receptor, with large negative values indicating higher selectivity for the target receptor, can be equivalently used to monitor selectivity.\cite{albanese2020structure} We will refer to the latter as the binding selectivity free energy (BSFE).

 Alchemical binding free energy computational models have been developed to estimate a ligand's standard binding free energy to a receptor or, more commonly in applied structure-based drug discovery, to estimate the relative binding free energy (RBFE) of a ligand pair to the same receptor.\cite{wang2015accurate,zou2019blinded,schindler2020large,lee2020alchemical,kuhn2020assessment,gapsys2020large,bieniek2021ties,hahn2022bestpractices,gapsys2022pre,cournia2017relative} RBFE models are useful for studying the relative potency of two inhibitors against the same receptor, but they do not provide information about the selectivity properties of either one. This work presents a receptor hopping alchemical protocol based on the Alchemical Transfer Method (ATM)\cite{azimi2022relative,sabanes2023validation,chen2024performance} that directly computes the BSFE of a ligand relative to two receptors.

 In binding free energy-based computational research for drug discovery, it is also common to integrate a set of calculated RBFE values using inference algorithms to estimate the binding free energies, or the absolute binding free energy (ABFE), of a set of ligands to one receptor.\cite{xu2019optimal,abel2017advancing} Here, we extend the DiffNet\cite{xu2019optimal} algorithm to obtain the binding free energies of a set of ligands to multiple receptors by integrating the results of RBFE and receptor hopping calculations. We also introduce a receptor swapping alchemical protocol that measures the free energy change for exchanging a pair of ligands across two receptors. We show that because the receptor swapping free energies are related to the differences of binding selectivity free energies, the DiffNet algorithm can estimate the BSFEs of a set of ligands against two receptors from the analysis of receptor swapping free energies.

 The primary aim of this work is to describe the computational algorithms and validate their correctness on simple molecular systems. We validate the alchemical free energy estimation protocols by computing the standard binding free energies of a set of guests to the TEMOA and TEETOA receptors,\cite{suating2020proximal,azimi2022application,amezcua2022overview} as well as the corresponding relative binding free energies, receptor hopping, and receptor swapping free energies by showing that they are consistent with each other. We also show how these quantities produce consistent standard binding free energy and binding selectivity free energies through DiffNet analysis. To anticipate future applications to medicinal systems, we illustrate the application of the receptor swapping methodology to model the relative binding selectivity of benzamidine and 1-amidinopiperidine to the trypsin and thrombin serine protease enzymes.

\section{Methods}

\subsection{Free Energies of Alchemical Transfer, Receptor Hopping, and Receptor Swapping}

Denote by $\Delta G^\circ_b(RL)$ the standard binding free energy between a receptor $R$ and a ligand $L$
\begin{equation} 
R +  L  \rightleftharpoons RL \quad\quad \Delta G^\circ_b(RL) \, .
\end{equation}
The relative binding free energy (RBFE), $\Delta G_r(R L_1, R L_2)$, of two ligands $L_1$ and $L_2$ to the a receptor $R$, corresponding to the equilibrium
\begin{equation} 
RL_1 + L_2  \rightleftharpoons RL_2 + L_1 \quad\quad \Delta G_r(R L_1, R L_2) \, ,
\end{equation}
and given by the difference of the standard binding free energies of ligands $L_2$ and $L_1$ to $R$
\begin{equation} 
\Delta G_r(R L_1, R L_2) =  \Delta G^\circ_b(RL_2) - \Delta G^\circ_b(RL_1)  \, ,
\end{equation}
is a measure of the relative affinity of the two ligands to the same receptor. To measure the relative affinity of one ligand for two receptors, $R_A$ and $R_B$, we define the binding selectivity free energy (BSFE) as the difference between the standard binding free energies of the two complexes 
\begin{equation} 
\Delta G_h(R_A L, R_B L) =  \Delta G^\circ_b(R_BL) - \Delta G^\circ_b(R_AL) \, ,
\label{eq:hop-abfe}
\end{equation} 
which corresponds to the process of transferring the ligand from one receptor to the other (hereafter referred to as receptor hopping)
\begin{equation} 
R_AL + R_B  \rightleftharpoons R_A + R_BL \quad\quad \Delta G_h(R_A L, R_B L) \, .
\end{equation}
The BSFE is related to the selectivity coefficient of ligand $L$ for receptor $R_B$ over $R_A$
\begin{equation} 
c(L) = \frac{K_b(R_BL)}{K_b(R_AL)}
\label{eq:sel-coeff-def}
\end{equation}
by the relation
\begin{equation} 
\Delta G_h(R_A L, R_B L) = -k_B T \ln \frac{K_b(R_BL)}{K_b(R_AL)} =  -k_B T \ln c(L) \, .
\label{eq:sel-coeff-from-dgh}
\end{equation}
Finally, let us consider the free energy of the process of swapping two ligands across two receptors
\begin{equation} 
R_A L_1 + R_B L_2  \rightleftharpoons R_A L_2 + R_B L_1 \quad\quad \Delta G_s(R_A L_1, R_B L_2) \, ,
\end{equation}
whose free energy, $\Delta G_s(R_A L_1, R_B L_2)$, can be written, by combining chemical equations, as either the difference of relative binding free energies or differences of binding selectivity free energies
\begin{eqnarray}
    \Delta G_s(R_A L_1, R_B L_2) & = & - \left[  \Delta G_r(R_B L_1, R_B L_2) - \Delta G_r(R_A L_1, R_A L_2)  \right]  \label{eq:swap-rbfe} \\
                                 & = & - \left[  \Delta G_h(R_A L_2, R_B L_2) - \Delta G_h(R_A L_1, R_B L_1)  \right]  \label{eq:swap-hop}
\end{eqnarray}
Eq.\ (\ref{eq:swap-rbfe}) is proven by combining the relative binding processes
\begin{align}
R_A L_1 + L_2  & \rightleftharpoons R_A L_2 + L_1 & \quad \Delta G_r(R_A L_1, R_A L_2)  \nonumber \\ 
R_B L_2 + L_1  & \rightleftharpoons R_B L_1 + L_2 & \quad -\Delta G_r(R_B L_1, R_B L_2) \nonumber \\  
\hline
R_A L_1 + R_B L_2 & \rightleftharpoons R_A L_2 + R_B L_1 & \quad  \Delta G_s(R_A L_1, R_B L_2)  \nonumber 
\end{align}
while Eq.\ (\ref{eq:swap-hop}) is obtained by combining the receptor hopping processes:
\begin{align}
R_A L_1 + R_B  & \rightleftharpoons R_A + R_B L_1 & \quad \Delta G_h(R_A L_1, R_B L_1)  \nonumber \\
R_B L_2 + R_A  & \rightleftharpoons R_B + R_A L_2 & \quad -\Delta G_h(R_A L_2, R_B L_2) \nonumber \\
\hline
R_A L_1 + R_B L_2 & \rightleftharpoons R_A L_2 + R_B L_1 & \quad  \Delta G_s(R_A L_1, R_B L_2) \nonumber 
\end{align}

Using Eqs.\ (\ref{eq:sel-coeff-from-dgh}) and (\ref{eq:swap-hop}),  the ratio of selectivity coefficients of two ligands for two receptors is related to the receptor swapping free energy by the relation
\begin{equation}
    \frac{c(L_2)}{c(L_1)} = e^{\Delta G_s(R_A L_1, R_B L_2)/k_B T}
    \label{eq:sel-coeff-ratio-from-RSFE}
\end{equation}

Each free energy change above (standard binding free energies, relative binding free energies, binding selectivity free energies, and receptor swapping free energies) corresponds to an alchemical transfer computational protocol. Standard binding free energies are obtained by the Absolute Binding Free Energy (ABFE) ATM protocol in which a ligand is transferred from the solution to the receptor binding site. Relative binding free energies are obtained by the Relative Binding Free Energy (RBFE) ATM protocol in which a ligand is transferred from the solution to the receptor binding site while the other ligand that is bound is simultaneously transferred in the opposite direction. Binding selectivity free energies are obtained by the Receptor Hopping Free Energy (RHFE) ATM protocol, in which the ligand is transferred from the binding site of the first receptor to the binding site of the second receptor. Finally, receptor swapping free energies are obtained using the receptor swapping Free Energy (RSFE) ATM protocol in which one ligand is transferred from one receptor to another while the other ligand is simultaneously transferred in the opposite direction.

\subsection{Diffnet Free Energy Estimation}

\begin{figure}
    \centering
    \includegraphics[scale = 0.45]{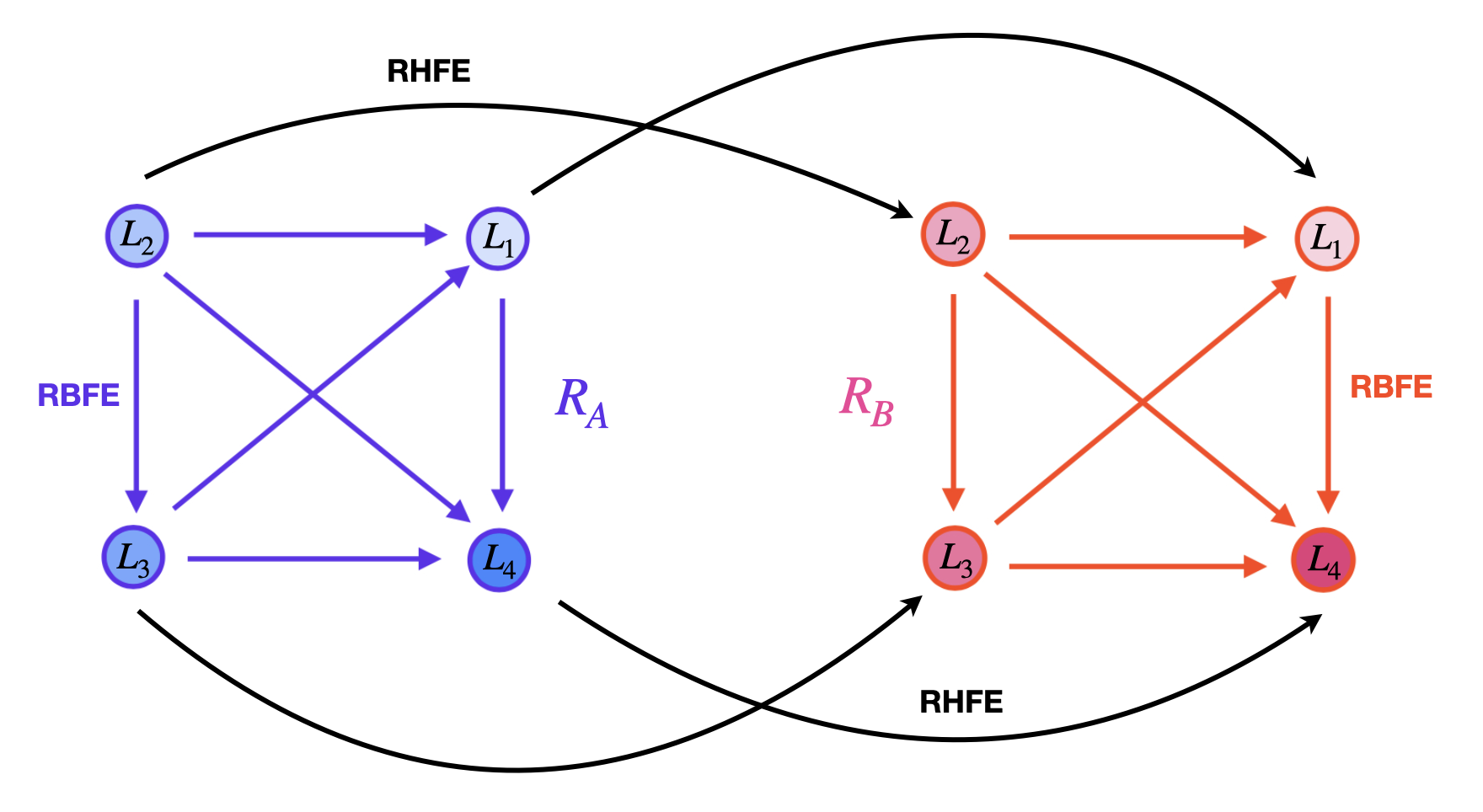}
    \caption{Network of alchemical transformations for the DiffNet analysis to obtain binding free energy estimates (ABFEs) for two receptors $R_A$ and $R_B$. Every circle represents a ligand in the RBFE scheme for a given receptor. Every colored arrow represents an RBFE calculation for a pair of ligands to a given receptor. Every black arrow represents a receptor hopping free energy of a ligand between two receptors.}
    \label{fig:diffnet-ABFE}
\end{figure}

DiffNet\cite{xu2019optimal} is a maximum likelihood procedure to find the standard binding free energies of a series of ligands to a receptor most consistent with a set of calculated relative binding free energies (RBFEs) given the standard binding free energy of a reference ligand. In this work, we extend DiffNet to yield the binding free energies of a set of ligands to two receptors given the binding free energy of one reference ligand to one receptor, the set of RBFEs against each receptor, and at least one receptor hopping free energy (RHFE) connecting the complex with one receptor to the corresponding complex with the other receptor (Figure \ref{fig:diffnet-ABFE}). DiffNet is based on minimizing the standard deviation-weighted squared differences between the differences of the target ABFEs and the input RBFEs. In the two-receptors extension proposed here, we add terms to the cost function of the form
\begin{equation}
    \frac{
    \left\{   \Delta G_h(R_A L, R_B L) -  [ \Delta G^\circ_b(R_BL) - \Delta G^\circ_b(R_AL) ]       \right\}^2
    }{
\sigma_h^2
    }
\end{equation}
from Eq.\ (\ref{eq:hop-abfe}) that enforces consistency between the differences between the variable standard binding free energies of each ligand for the two receptors and the corresponding calculated receptor hopping free energy values, $\Delta G_h(R_A L, R_B L)$, weighted by its statistical variance $\sigma^2_h$.  These terms connect the network of binding free energy differences of one receptor to the other to ensure that the resulting binding free energies are on the same scale and reflect the ligands' selectivity for the two receptors. The scheme can be generalized to any number of receptors as long as sufficient receptor hopping free energies connect each receptor's sub-networks.

\begin{figure}
    \centering
    \includegraphics[scale=0.45]{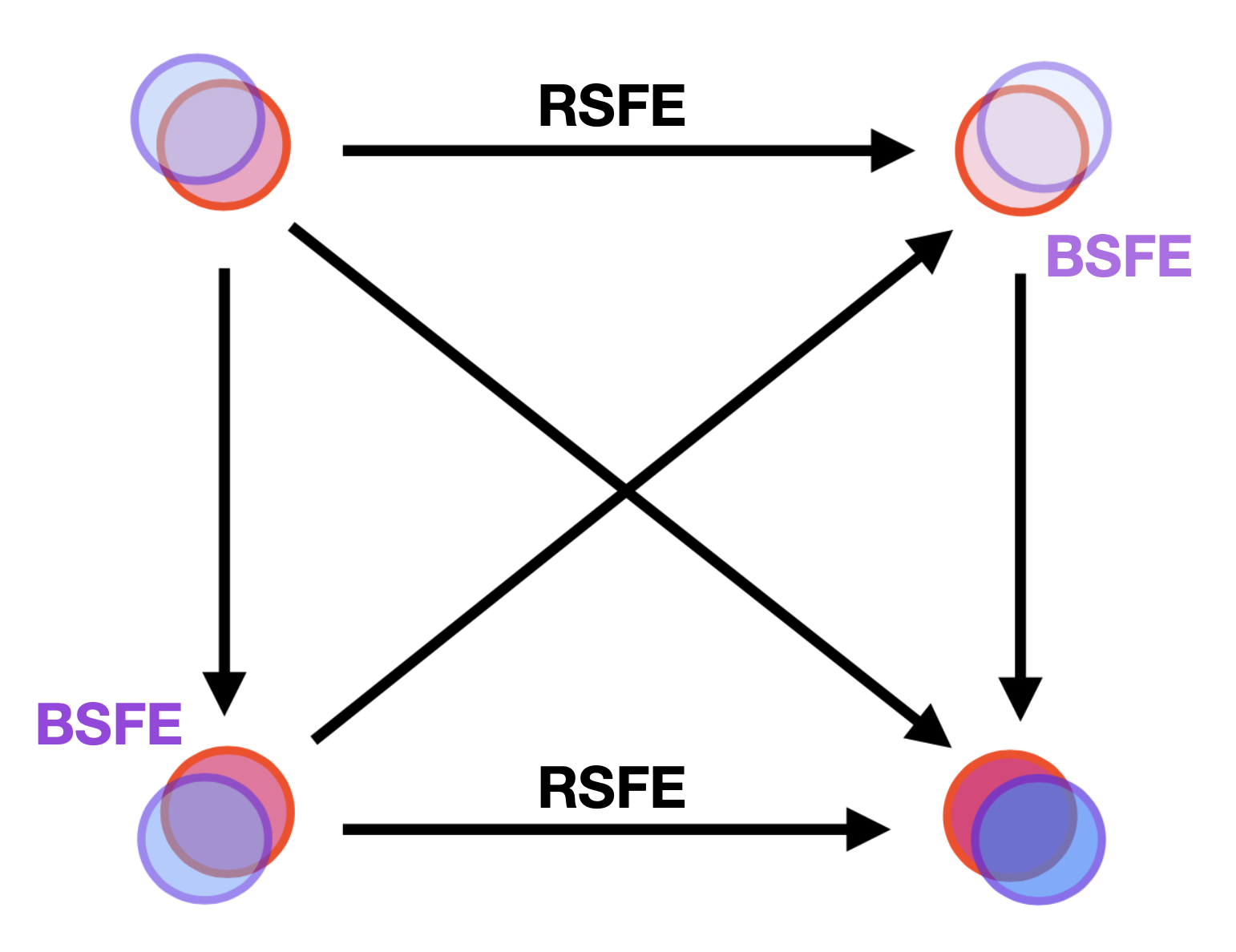}
    \caption{Network of alchemical transformations for the DiffNet analysis to obtain the binding selectivity free energies of a set of ligands for a pair of receptor. Each circle pair represents a binding selectivity free energy (BSFE) of a ligand for the two receptors, and each arrow represents a receptor swapping free energy (RSFE) estimate for two ligands.}
    \label{fig:diffnet-HFE}
\end{figure}

In cases where only the selectivity of the ligands for two receptors is of interest rather than the strength of the individual affinities, we consider the network of free energy differences pictured in Figure \ref{fig:diffnet-HFE} where the nodes, represented by overlapping circles, represent the binding selectivity free energies that we seek and the directed edges represent their differences estimated by receptor swapping calculations. Formally, this scenario is equivalent to the standard DiffNet strategy to estimate binding free energies from RBFEs estimates, where the binding selectivity free energies (BSFEs) replace the binding free energies and the RBFEs are replaced by receptor swapping free energies (RSFEs). In both cases, the network nodes are the quantities we estimate, and the edges represent their differences. To accomplish selectivity analysis, the DiffNet cost function includes terms that restrain the differences of BSFEs to the differences of RSFEs according to Eq.\ (\ref{eq:swap-hop}) 
\begin{equation}
    \frac{
    \left\{   \Delta G_s(R_A L_1, R_B L_2) -  [ - \Delta G_h(R_A L_2, R_B L_2) + \Delta G_h(R_A L_1, R_B L_1) ]       \right\}^2
    }{
\sigma_s^2
    }
\end{equation}
where $\sigma^2_s$ is the statistical variance of the RSFE estimate. In this way, the BSFEs of a set of ligands for two receptors are obtained from a set of RSFEs and the BSFE of one reference ligand for the two receptors.

\section{Computational Details}

\subsection{System Setup and Simulation Settings}

\begin{figure}
    \centering
    \includegraphics[scale=1.5]{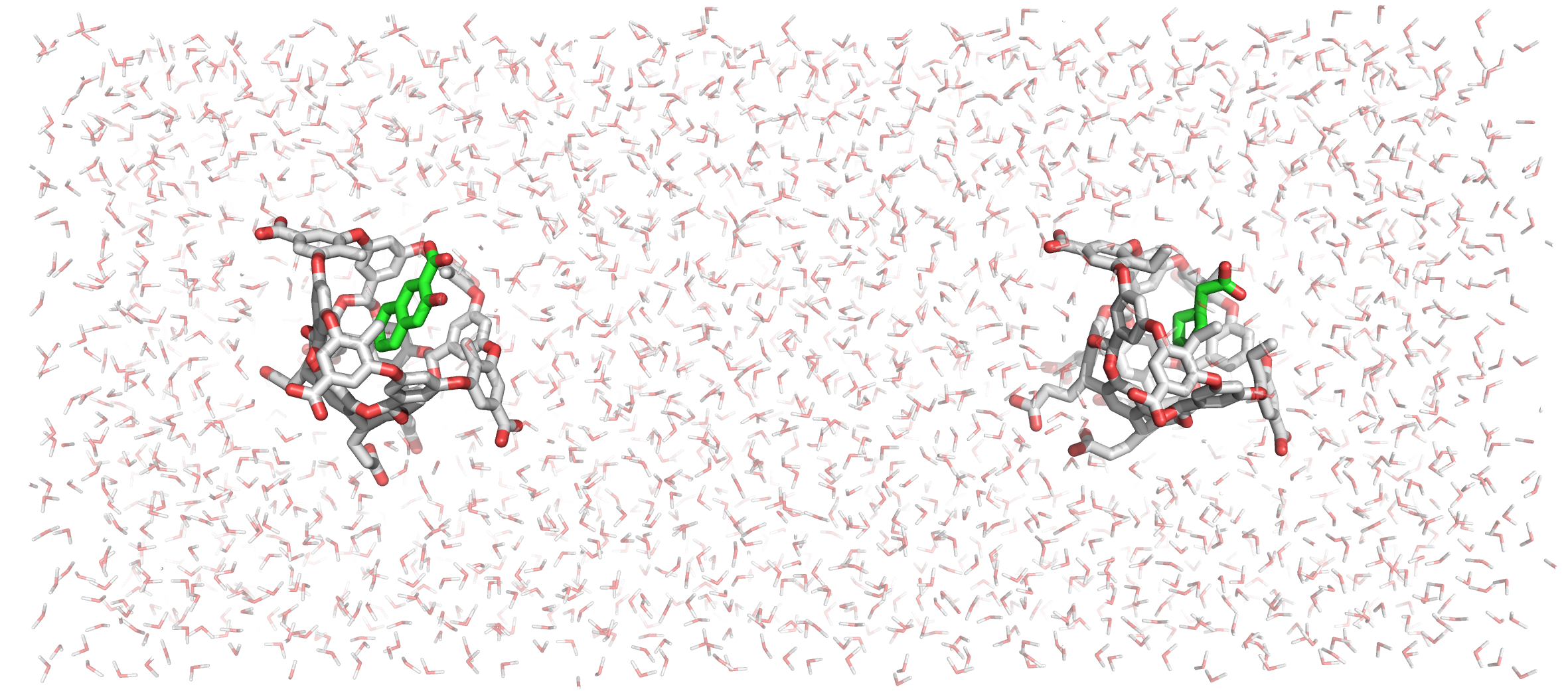}
    \caption{\label{fig:sim-box} The simulation system of the host-guest complexes investigated in this work. In ABFE calculations, a ligand (green carbon atoms) is transferred from one host to the solvent region at the center of the simulation box in between the two hosts (gray carbon atoms). In RBFE calculations, the first ligand of the pair is transferred from a host to the center of the box, while the second ligand is translated from the center of the box to the host. In receptor hopping free energy (RHFE) calculations, a ligand is transferred from one host to the other. In receptor swapping free energy (RSFE) calculations, one ligand is transferred from its bound host to the other host while a second ligand is translated in the opposite direction. }   
\end{figure}

The host-guest alchemical calculations employed the setup illustrated in Fig.\ \ref{fig:sim-box}. The TEMOA and TEETOA hosts were placed 50 \AA\ apart along the x-direction. They were kept at their positions using flat-bottom harmonic positional restraints with a tolerance of $0.5$ \AA\ of and a force constant of $25$ kcal/(mol \AA$^2$) on the atoms of the lower portion of the cup of each host as in reference \citenum{azimi2022application}. In the ATM absolute binding free energy calculations, the guest was displaced from the binding site of the host by $25$ \AA\ in either the positive or negative direction to place it at the midpoint position in the solvent between the two hosts. The same strategy was used for the relative binding free energy calculations (RBFE) for each host, except that one guest was displaced from the binding site to the solvent while the other guest was simultaneously displaced in the opposite direction. In the receptor hopping calculations, the guest was displaced by $50$ \AA\ along the x-axis to move it from the binding site of TEMOA to that of TEETOA. The receptor swapping calculations were implemented in the same way, except that one guest was displaced from TEMOA to TEETOA while the other guest was simultaneously displaced in the opposite direction.

\begin{figure}
    \centering
    \includegraphics[scale=0.35]{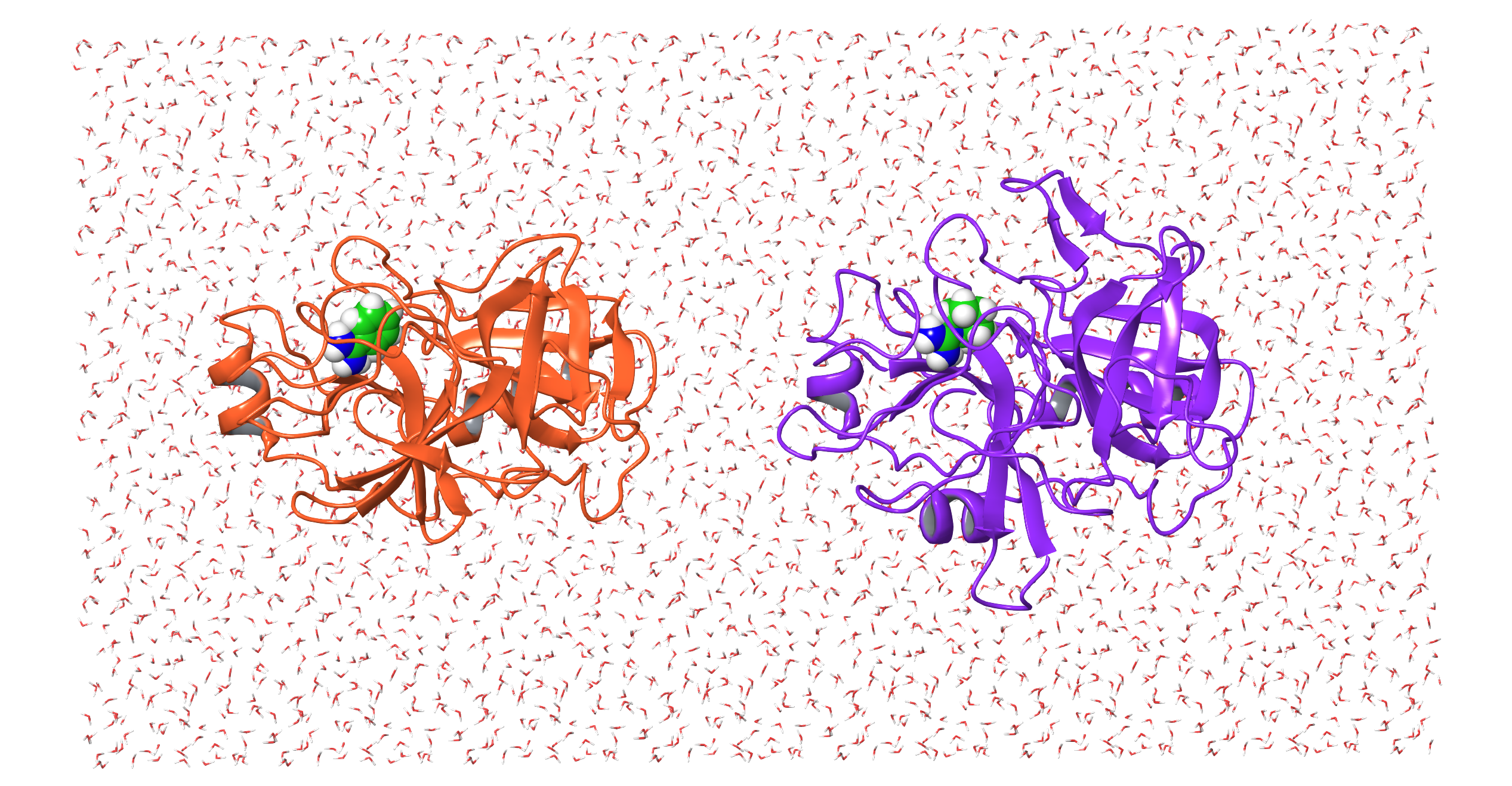}
    \caption{\label{fig:RSFE-protein-ligand} The simulation system of the protein-ligand complexes investigated in this work. The proteins are shown in ribbon representation; trypsin (left) is orange and thrombin (right) is purple. The ligands are shown in Van der Waals representation: benzamidine is on the left bound to trypsin and 1-amidinopiperidine is on the right bound to thrombin. The two complexes are displaced by 60 \AA\ along the z-axis, which is arranged horizontally and pointing from left to right.}   
\end{figure}

The system for the RBFE and RSFE calculations of the protein-ligand complexes was arranged similarly. Trypsin and thrombin were aligned and placed 60 \AA\ apart along the z-direction with benzamidine bound to trypsin and amidinopiperidine bound to thrombin, initially (Figure \ref{fig:RSFE-protein-ligand}). To maintain their relative positions, the C-$\alpha$ atoms of each receptor were positionally restrained with a flat-bottom harmonic potential with a tolerance of 1.5 \AA\ and a force constant of 25 kcal/(mol \AA$^2$). During the RSFE alchemical calculations, benzamidine was translated into the thrombin's binding, and simultaneously, amidinopiperidine was displaced into trypsin's binding site. The RBFE calculations of the two ligands were implemented similarly to the host-guest systems described above. AToM-OpenMM input files with a complete set of simulation settings are available at \url{https://github.com/Gallicchio-Lab/receptor-hopping}.

The guests and the ligands were docked into the respective receptor binding sites hosts using Maestro (Schr\"{o}dinger, Inc.). A flat-bottom harmonic positional restraint with a force constant $k_c = 25$ kcal/mol/\AA$^2$ and tolerance of $4.5$ \AA\ on the centers of mass of the ligand and the receptor as in reference \citenum{azimi2022application} was applied to define the binding site region.\cite{Gilson:Given:Bush:McCammon:97,Gallicchio2011adv} In the RBFE calculations, alignment restraints\cite{azimi2022relative} were applied to each pair of guests using the reference atoms and forced constants as specified in the input files at \url{https://github.com/Gallicchio-Lab/receptor-hopping}. Guest and ligands were assigned GAFF2 parameters, and the protein receptors were parameterized using the Amber ff14SB force field.\cite{maier2015ff14sb} The receptors and ligands were combined on tLeaP. The systems were minimized and thermalized before production calculations. 

The host-guest alchemical replica exchange molecular dynamics simulations were conducted with the AToM-OpenMM software version 3.5.0\cite{AToM-OpenMM-23} with OpenMM 7.7\cite{eastman2017openmm} and the ATM Meta Force plugin version 0.3.5\cite{ATMMetaForce-OpenMM-plugin}. The protein-ligand calculations were performed later during the project using the newer 8.1.1 versions of AToM-OpenMM and OpenMM.\cite{AToM-OpenMM-24,eastman2023openmm} The alchemical calculations employed 22 alchemical states with the non-linear softplus alchemical potential function\cite{pal2019perturbation,khuttan2021alchemical} and the alchemical schedules provided in the input files repository at \url{https://github.com/Gallicchio-Lab/receptor-hopping}. The soft-core perturbation energy function with parameters $u_{\rm max} = 200$ kcal/mol, $u_c = 100$ kcal/mol, and $a=1/16$ was used.\cite{khuttan2021alchemical} Molecular dynamics with an MD time-step of 2 fs was conducted for 40 ns per replica for the host-guest calculations and 20 ns per replica for the protein-ligand calculations. Replicas executions were cycled every 40 ps on GPU devices according to the asynchronous replica exchange algorithm.\cite{Gallicchio2008,gallicchio2015asynchronous,xia2015large} Temperature was maintained at 300 K using the Langevin thermostat with a time constant of $0.5$ ps. Perturbation energy samples were collected every 40 ps, and free energies were estimated using the UWHAM method\cite{Tan2012,UWHAM-in-R-2013} after discarding 1/3 of the earliest portion of the trajectory. See \url{https://github.com/Gallicchio-Lab/receptor-hopping} to access detailed simulation settings and free energy analysis scripts. Generalized DiffNet analysis was performed using the \texttt{diffnet-tf} package available at \url{https://github.com/Gallicchio-Lab/diffnet-tf}.

\section{Results}

\subsection{Host-Guest Systems}

\begin{figure}
    \centering
    \includegraphics[scale = 0.5]{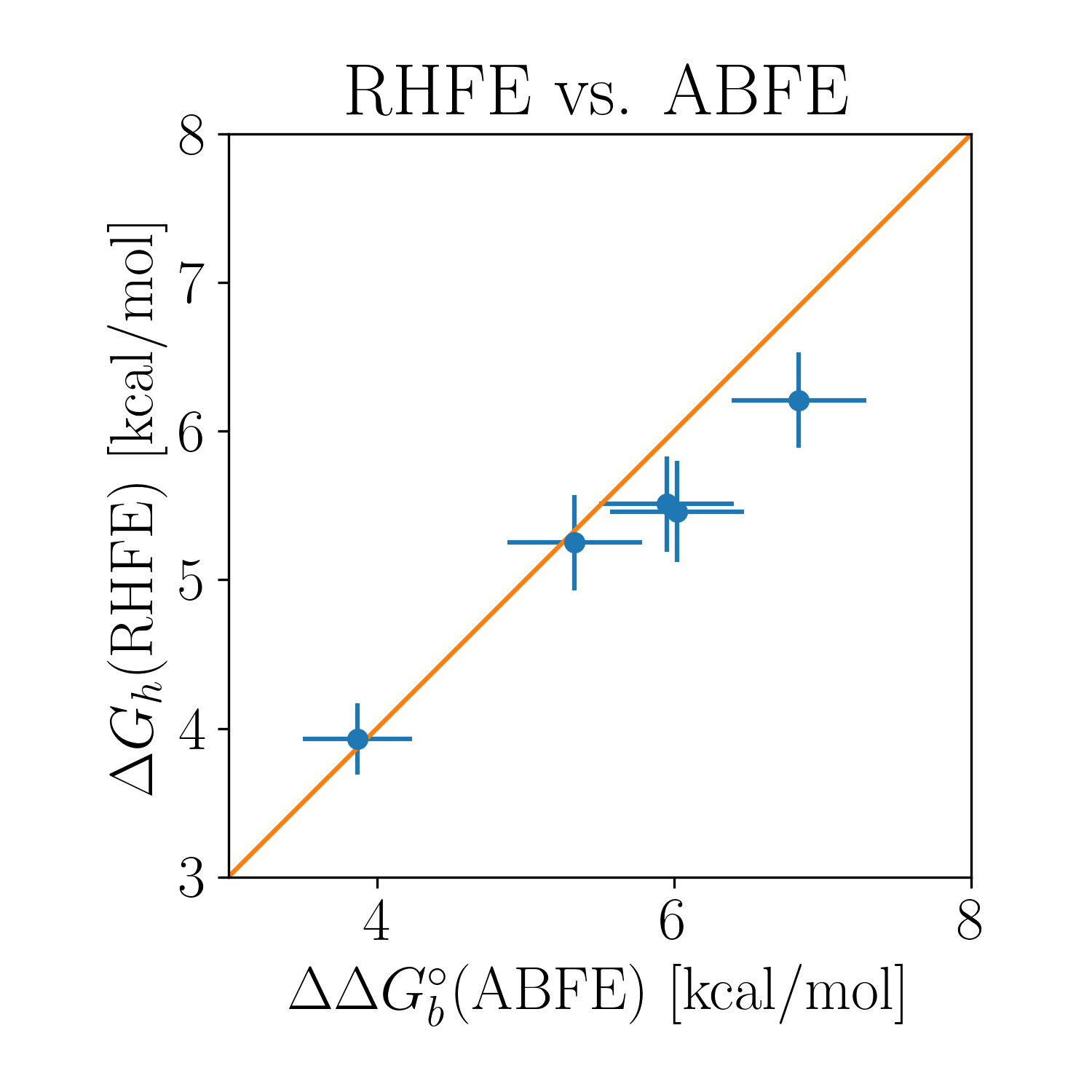}
    \caption{Scatterplot of the receptor hopping free energy estimates with respect to the differences of absolute binding free energies. The diagonal line corresponds to perfect agreement.}
    \label{fig:dg-hop}
\end{figure}

The standard binding free energies of the five guests calculated using the ABFE protocol and simulation system of Figure \ref{fig:sim-box} are listed in Table \ref{tab:dg-absolute}. These ABFE estimates are consistent with the experiments and the estimates we reported earlier as part of the SAMPL8 challenge,\cite{azimi2022application} confirming that the guests bind more strongly to the TEMOA host than the TEETOA host. The free energies for transferring each guest from the TEMOA to the TEETOA hosts (receptor hopping) are presented in the second column of Table \ref{tab:dg-hopping}. Receptor hopping free energies are equivalent to the differences of absolute binding free energies (ABFEs) listed in the third column of Table \ref{tab:dg-hopping}. The Root Mean Square Deviation (RMSD) between the two sets of receptor hopping free energy estimates is small, and the corresponding values are within statistical uncertainty based on the two-sided t-test with a p-value confidence level of 5\%. The uncertainties of the estimates from the differences ABFEs are consistently larger than those from receptor hopping by 35\% or more. The correspondence between these quantities is also presented in Figure \ref{fig:dg-hop}.

\begin{figure}
    \centering
    \includegraphics[scale = 0.4]{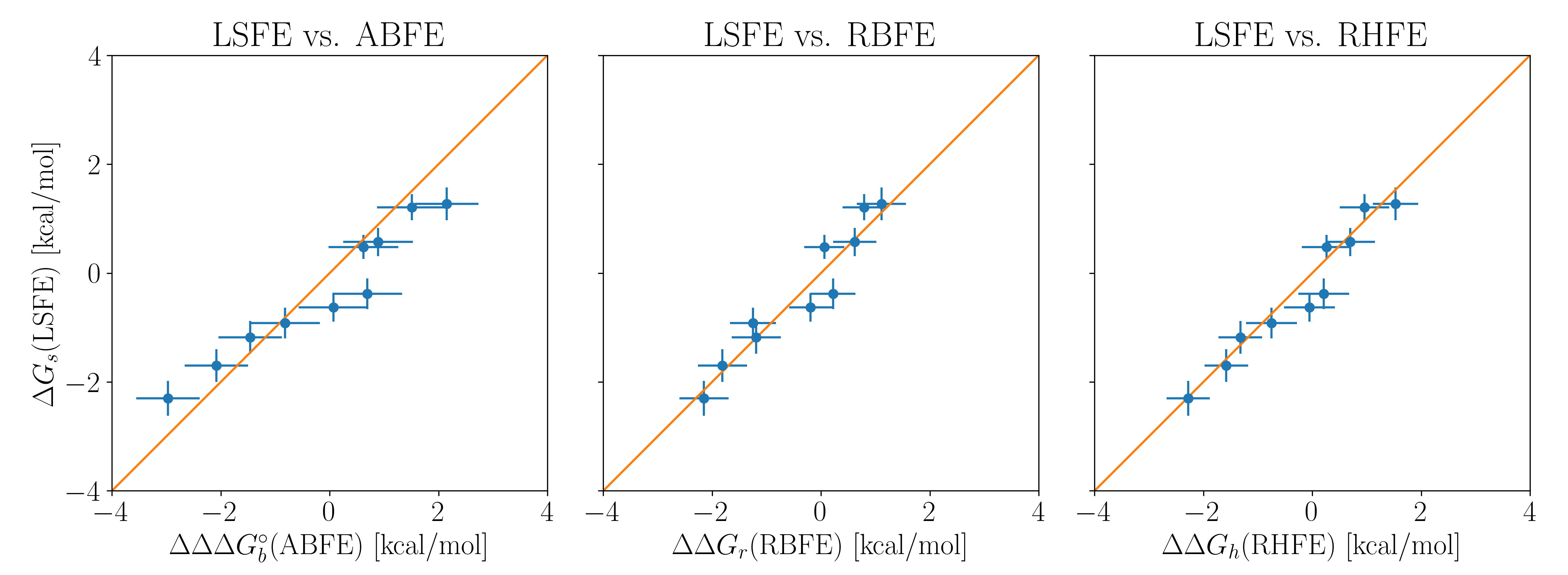}
    \caption{Scatterplot of the receptor swapping free energy estimates with respect to the differences of absolute binding free energies (left), relative binding free energies (middle), and receptor hopping free energies (right). The diagonal line corresponds to perfect agreement.}
    \label{fig:dg-swap}
\end{figure}

Next, the calculated relative binding free energies (RBFEs) to the TEMOA and TEETOA hosts are reported in Table \ref{tab:dg-relative} for all pairs of guests. Again, these are compared to the corresponding differences of the ABFEs from Table \ref{tab:dg-absolute}, generally observing agreement within statistical uncertainty. The only cases where the deviation is large enough to have less than a 5\% probability of having occurred by chance are the G1/G2p and the G2p/G4 pairs of the TEETOA host. Finally, the calculated free energies for swapping all pairs of guests across the two hosts (receptor swapping) are reported in Table \ref{tab:dg-swapping} and Figure \ref{fig:dg-swap} compared to the corresponding estimates from the differences ABFEs, RBFEs, and receptor hopping free energies [Eqs.\ \ref{eq:swap-rbfe} and \ref{eq:swap-hop}]. We observe an excellent agreement between these quantities, with RMSD values well within statistical uncertainties, supporting the correctness of the implementation of the receptor hopping and receptor swapping protocols.

\begin{table}
\caption{\label{tab:dg-absolute} ATM ABFE estimates of the standard binding free energies of the SAMPL8 guests to the TEMOA and TEETOA hosts.}
\begin{center}
\sisetup{separate-uncertainty}
\begin{tabular}{l S[table-format = 3.2(2)] S[table-format = 3.2(2)] }
 Guest & \multicolumn{1}{c}{$\Delta G^\circ_b$$^{a,b}$} & \multicolumn{1}{c}{$\Delta G^\circ_b$$^{a,b}$} \\ [0.5ex]
\hline
     & \multicolumn{1}{c}{TEMOA} & \multicolumn{1}{c}{TEETOA} \\
G1   &  -6.65(32) & -0.63(32) \\ 
G2p  & -12.10(26) & -8.23(26) \\ 
G3   &  -8.80(32) & -1.96(32) \\ 
G4   &  -8.18(32) & -2.23(32) \\ 
G5   &  -7.97(32) & -2.64(32) \\ 
\hline
\end{tabular}
\begin{flushleft}\small
$^a$In kcal/mol. $^b$Uncertainties are reported as twice the standard deviation.
\end{flushleft}
\end{center}
\end{table}

\begin{table}
\caption{\label{tab:dg-hopping} ATM estimates of the receptor hopping free energies (RHFEs) of the SAMPL8 guests from the TEMOA to the TEETOA hosts compared to corresponding differences of ABFEs from Table \ref{tab:dg-absolute}.}
\begin{center}
\sisetup{separate-uncertainty}
\begin{tabular}{l S[table-format = 3.2(2)] S[table-format = 3.2(2)] }
 Guest & \multicolumn{1}{c}{$\Delta G_h$$^{a,b}$} & \multicolumn{1}{c}{$\Delta G_h$$^{a,b,c}$} \\ [0.5ex]
      & \multicolumn{1}{c}{(RHFE)} & \multicolumn{1}{c}{(from ABFEs)}                                               \\ [0.5ex]
\hline
G1  &  5.46(34) &  6.03(46)  \\ 
G2p &  3.93(24) &  3.87(38)  \\
G3  &  6.21(32) &  6.83(46)  \\ 
G4  &  5.51(32) &  5.95(46)  \\
G5  &  5.25(32) &  5.33(46)  \\
\hline
\multicolumn{1}{c}{RMSD$^d$}  &    &  \multicolumn{1}{c}{$0.43$} \\
\hline
\end{tabular}
\begin{flushleft}\small
$^a$In kcal/mol. $^b$Uncertainties are reported as twice the standard deviation. $^c$ From Table \ref{tab:dg-absolute}. $^d$ Root mean square deviation in kcal/mol.
\end{flushleft}
\end{center}
\end{table}

\begin{table}
\caption{\label{tab:dg-relative} ATM estimates of the relative binding free energies (RBFEs) of the SAMPL8 guests to the TEMOA and TEETOA hosts compared to corresponding differences of ABFEs from Table \ref{tab:dg-absolute} .}
\begin{center}
\sisetup{separate-uncertainty}
\begin{tabular}{l S[table-format = 3.2(2)] S[table-format = 3.2(2)] S[table-format = 3.2(2)] S[table-format = 3.2(2)]}
 Pair & \multicolumn{1}{c}{$\Delta G_r$$^{a,b}$} & \multicolumn{1}{c}{$\Delta G_r$$^{a,b,c}$} &
        \multicolumn{1}{c}{$\Delta G_r$$^{a,b}$} & \multicolumn{1}{c}{$\Delta G_r$$^{a,b,c}$} \\ [0.5ex]
      & \multicolumn{1}{c}{(RBFE)} & \multicolumn{1}{c}{(from ABFEs)} &
        \multicolumn{1}{c}{(RBFE)} & \multicolumn{1}{c}{(from ABFEs)} \\ [0.5ex]
        
\hline
      & \multicolumn{2}{c}{TEMOA} & \multicolumn{2}{c}{TEETOA} \\
G1 to G2p &  -5.08(32) & -5.45(42) &  -6.19(32) & -7.61(42) \\ 
G1 to G3  &  -1.73(30) & -2.15(46) &  -0.48(30) & -1.34(46) \\
G1 to G4  &  -1.85(28) & -1.53(46) &  -1.66(28) & -1.61(46) \\ 
G1 to G5  &  -1.40(28) & -1.31(46) &  -1.62(30) & -2.01(46) \\
G2p to G3 &   3.30(32) &  3.31(42) &   5.45(32) &  6.27(42) \\
G2p to G4 &   3.05(32) &  3.93(42) &   4.86(32) &  6.00(42) \\
G2p to G5 &   3.50(32) &  4.14(42) &   4.69(32) &  5.60(42) \\
G3  to G4 &  -0.12(28) &  0.62(46) &  -0.74(28) & -0.27(46) \\
G3  to G5 &  -0.04(28) &  0.83(46) &  -0.83(28) & -0.67(42) \\
G4  to G5 &   0.11(26) &  0.21(46) &   0.05(26) & -0.46(46) \\ 
\hline
\multicolumn{1}{c}{RMSD$^d$}  &    & \multicolumn{1}{c}{$0.54$} &  & \multicolumn{1}{c}{$0.78$} \\
\hline
\end{tabular}
\begin{flushleft}\small
$^a$In kcal/mol. $^b$Uncertainties are reported as twice the standard deviation. $^c$ From Table \ref{tab:dg-absolute}. $^d$Relative to calculated RBFEs.
\end{flushleft}
\end{center}
\end{table}

\begin{table}
\caption{\label{tab:dg-swapping} ATM estimates of the receptor swapping free energies (RSFEs) of pairs of guests between the TEMOA and TEETOA hosts compared to corresponding differences of absolute binding free energies (ABFEs), relative binding free energies (RBFEs), and receptor hopping free energies (RHFEs) from Tables \ref{tab:dg-absolute},  \ref{tab:dg-relative}, and \ref{tab:dg-hopping}, respectively.}
\begin{center}
\sisetup{separate-uncertainty}
\begin{tabular}{l S[table-format = 3.2(2)] S[table-format = 3.2(2)] S[table-format = 3.2(2)] S[table-format = 3.2(2)]}
 Pair & \multicolumn{1}{c}{$\Delta G_s$$^{a,b}$}   & \multicolumn{1}{c}{$\Delta G_s$$^{a,b,c}$} &
        \multicolumn{1}{c}{$\Delta G_s$$^{a,b,d}$} & \multicolumn{1}{c}{$\Delta G_s$$^{a,b,e}$} \\ [0.5ex]
      & \multicolumn{1}{c}{(RSFE)} & \multicolumn{1}{c}{(from ABFEs)} &
        \multicolumn{1}{c}{(from RBFEs)} & \multicolumn{1}{c}{(from RHFEs)} \\ [0.5ex]
\hline
G1, G2p &   1.27(30) &  2.16(58) &   1.10(45) &  1.53(42) \\ 
G1, G3  &  -0.92(28) & -0.81(64) &  -1.25(42) & -0.75(47) \\
G1, G4  &  -0.63(26) &  0.08(64) &  -0.19(40) & -0.06(47) \\ 
G1, G5  &  -0.38(28) &  0.70(64) &   0.21(41) &  0.21(47) \\
G2p, G3 &  -2.30(32) & -2.96(58) &  -2.14(45) & -2.28(40) \\
G2p, G4 &  -1.70(30) & -2.08(58) &  -1.81(45) & -1.59(40) \\
G2p, G5 &  -1.18(30) & -1.46(58) &  -1.19(45) & -1.32(40) \\
G3, G4  &   0.57(26) &  0.89(64) &   0.62(40) &  0.69(45) \\
G3, G5  &   1.21(24) &  1.50(64) &   0.80(40) &  0.96(45) \\
G4, G5  &   0.48(22) &  0.61(64) &   0.06(37) &  0.26(45) \\ 
\hline
\multicolumn{1}{c}{RMSD$^f$}  &    & \multicolumn{1}{c}{$0.57$} &  \multicolumn{1}{c}{$0.33$}  &  \multicolumn{1}{c}{$0.31$} \\
\hline
\end{tabular}
\begin{flushleft}\small
$^a$In kcal/mol. $^b$Uncertainties are reported as twice the standard deviation. $^c$ From Table \ref{tab:dg-absolute}. $^d$ From Table \ref{tab:dg-relative}. $^e$ From Table \ref{tab:dg-hopping}.   $^f$Relative to calculated RBFEs.
\end{flushleft}
\end{center}
\end{table}

Table \ref{tab:dg-diffnet-fromhopping} reports the binding free energy estimates of the guests to both hosts obtained from the DiffNet analysis of the receptor hopping free energies (RHFEs) in Table \ref{tab:dg-hopping} and the relative binding free energies (RBFEs) from Table \ref{tab:dg-relative}, using only the binding free energy of G1 to TEMOA as a reference. DiffNet is commonly used to analyze RBFE data for individual receptors, and its results cannot be used to infer receptor selectivities. In contrast, this experiment illustrates the scenario in which we estimate a consistent set of binding free energies of a series of ligands to two receptors using RBFE data for one receptor and RHFE data from one receptor to the other, using only the complex with one receptor as the reference (Figure \ref{fig:diffnet-ABFE}). The differences between the binding free energy estimates for the two receptors obtained in this way reflect the selectivity binding free energies of the ligands. The estimates in the second column of Table \ref{tab:dg-diffnet-fromhopping} result from including the RHFEs from TEMOA to TEETOA of all guests. The values in the third column result from the inclusion of only the RHFE of G1. In either case, as can be seen from the small RMSD, the Diffnet estimates of the binding free energies are in good agreement with the corresponding direct ABFE estimates. 

\begin{table}
\caption{\label{tab:dg-diffnet-fromhopping} Diffnet estimates of the standard binding free energies of the SAMPL8 guests to the TEMOA and TEETOA hosts from the receptor hopping (RHFE) and relative binding (RBFE) free energy estimates, compared to the calculated absolute binding free energies (ABFE) from Table \ref{tab:dg-absolute}.}
\begin{center}
\sisetup{separate-uncertainty}
\begin{tabular}{l S[table-format = 3.2(2)] S[table-format = 3.2(2)] S[table-format = 3.2(2)] }
 Guest & \multicolumn{1}{c}{$\Delta G^\circ_b$$^{a,b}$} & \multicolumn{1}{c}{$\Delta G^\circ_b$$^{a,b}$} & \multicolumn{1}{c}{$\Delta G^\circ_b$$^{a,b,c}$} \\ [0.5ex]
       & \multicolumn{1}{c}{(Diffnet, from all RHFEs)} & \multicolumn{1}{c}{(Diffnet, G1's RHFE only)} & \multicolumn{1}{c}{(ABFE)}  \\ [0.5ex]
\hline
      \multicolumn{4}{c}{TEMOA} \\
G1$^\ast$   &  \multicolumn{1}{c}{-6.65} & \multicolumn{1}{c}{-6.65} &  -6.65(32) \\ 
G2p   &  -11.60(13)                &  -11.64(20)               & -12.10(26) \\ 
G3    &   -8.28(13)                &   -8.31(20)               &  -8.80(32) \\ 
G4    &   -8.42(12)                &   -8.45(16)               &  -8.18(32) \\ 
G5    &   -8.20(10)                &   -8.23(21)               &  -7.97(32) \\ 
\hline
\multicolumn{1}{c}{RMSD$^d$}  &    & \multicolumn{1}{c}{$0.03$} &  \multicolumn{1}{c}{$0.40$} \\
\hline
      \multicolumn{4}{c}{TEETOA} \\
G1    &   -1.35(20)                &   -1.19(38)               &  -0.63(32) \\ 
G2p   &   -7.62(20)                &   -7.42(42)               &  -8.23(26) \\ 
G3    &   -2.06(17)                &   -1.88(43)               &  -1.96(32) \\ 
G4    &   -2.90(18)                &   -2.71(44)               &  -2.23(32) \\ 
G5    &   -2.91(10)                &   -2.72(40)               &  -2.64(32) \\
\hline
\multicolumn{1}{c}{RMSD$^d$}  &    & \multicolumn{1}{c}{$0.19$} &  \multicolumn{1}{c}{$0.53$} \\
\hline
\end{tabular}
\begin{flushleft}\small
$^a$In kcal/mol. $^b$Uncertainties are reported as twice the standard deviation. $^c$From Table \ref{tab:dg-absolute}. $^d$Relative to DiffNet ABFE values from all RHFEs. $^\ast$Reference complex.
\end{flushleft}
\end{center}
\end{table}

\begin{figure}
    \centering
    \includegraphics[scale = 0.6]{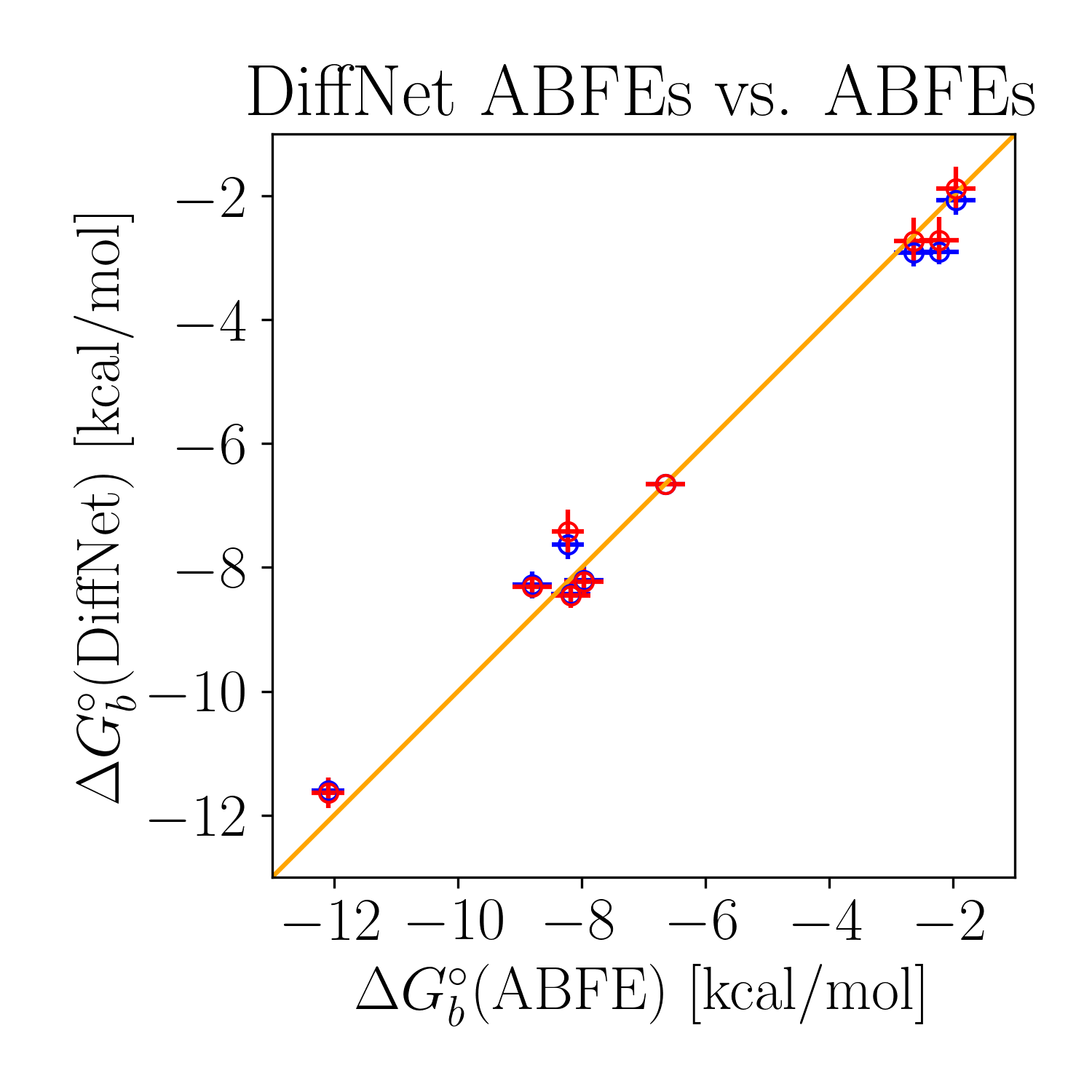}
    \caption{Scatterplot of the DiffNet-estimated standard binding free energies of the SAMPL8 complexes with respect to the calculated ABFEs from Table \ref{tab:dg-absolute}. Blue markers denote DiffNet estimates using all calculated RHFEs, and red markers denote those from only the RHFE of the G1 guest. The diagonal line corresponds to perfect agreement.}
    \label{fig:dg-diffnet-fromhopping}
\end{figure}

The Diffnet algorithm estimates of the binding selectivity free energies using the receptor swapping data in Table \ref{tab:dg-swapping} and setting the calculated RHFE of G1 as the reference binding selectivity free energy are shown in Table \ref{tab:dg-diffnet-fromsbfes}. The Diffnet binding selectivity free energy estimates are in good agreement with the direct RHFE calculations from Table \ref{tab:dg-hopping}. This experiment illustrates the scenario in which we measure the binding selectivity of a series of ligands for two receptors using receptor swapping free energy data of ligand pairs across the two receptors, knowing the selectivity binding free energy of at least one ligand (Figure \ref{fig:diffnet-HFE}).

\begin{table}
\caption{\label{tab:dg-diffnet-fromsbfes} Diffnet estimates of the selectivity binding free energies of the SAMPL8 guests between the TEMOA and TEETOA hosts from the receptor swapping free energy (RSFE) estimates, compared to the calculated receptor hopping free energies (RHFEs) from Table \ref{tab:dg-hopping}.}
\begin{center}
\sisetup{separate-uncertainty}
\begin{tabular}{l S[table-format = 3.2(2)] S[table-format = 3.2(2)] }
 Guest & \multicolumn{1}{c}{$\Delta G_h$$^{a,b}$} & \multicolumn{1}{c}{$\Delta G_h$$^{a,b,c}$} \\ [0.5ex]
       & \multicolumn{1}{c}{(Diffnet, from RSFEs)} & \multicolumn{1}{c}{(RHFE)}  \\ [0.5ex]
\hline
G1$^\ast$   &  \multicolumn{1}{c}{5.46}  &   5.46(34)  \\ 
G2p   &  4.31(20)                  &   3.93(24)  \\ 
G3    &  6.61(13)                  &   6.21(32)  \\ 
G4    &  6.05(18)                  &   5.51(32)  \\ 
G5    &  5.56(20)                  &   5.25(32)  \\ 
\hline
\multicolumn{1}{c}{RMSD}  &    & \multicolumn{1}{c}{$0.41$} \\
\hline
\end{tabular}
\begin{flushleft}\small
$^a$In kcal/mol. $^b$Uncertainties are reported as twice the standard deviation. $^c$From Table \ref{tab:dg-hopping}. $^\ast$Reference.
\end{flushleft}
\end{center}
\end{table}

\begin{figure}
    \centering
    \includegraphics[scale = 0.6]{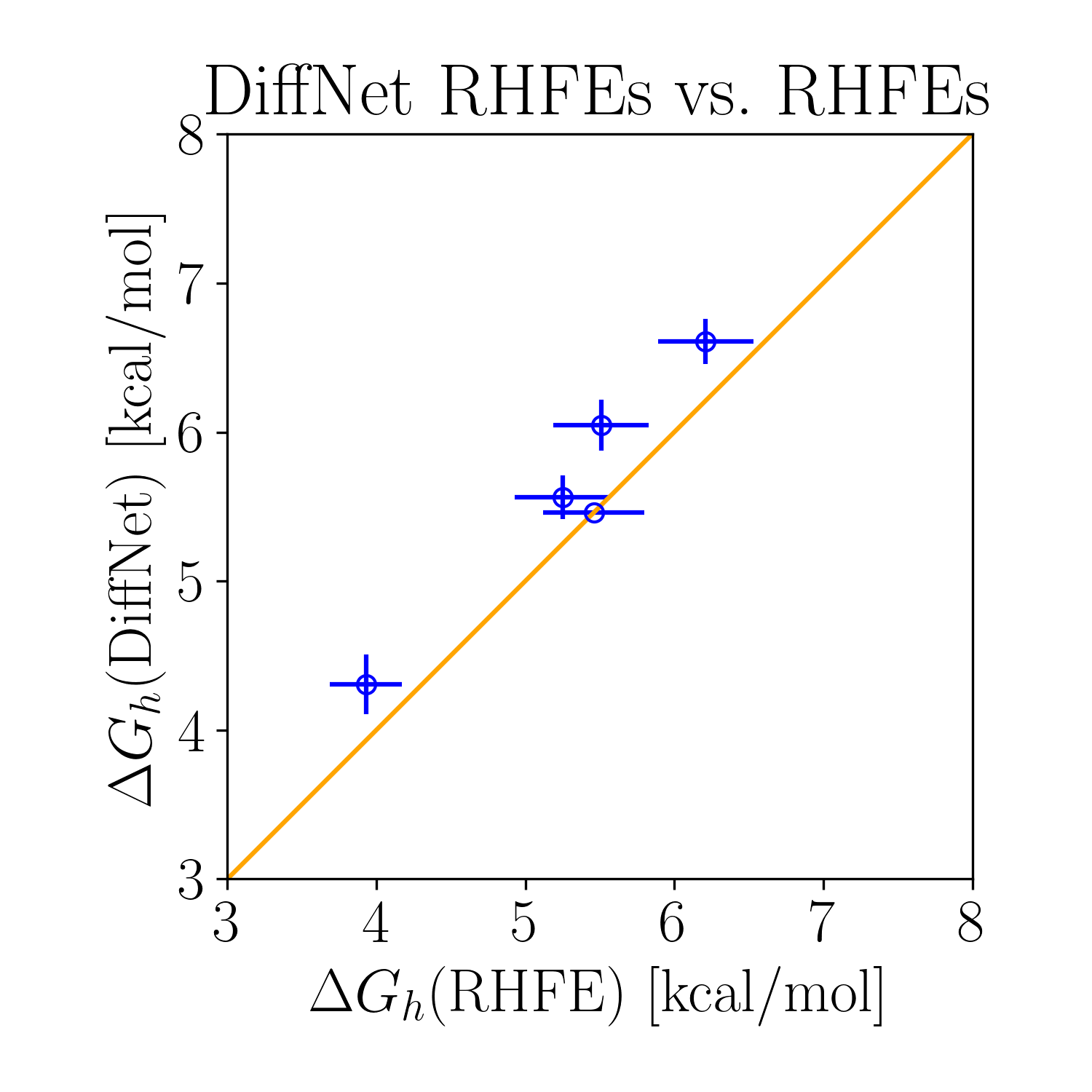}
    \caption{Scatterplot of the DiffNet-estimated selectivity binding free energies of the SAMPL8 complexes with respect to the calculated RHFEs from Table \ref{tab:dg-hopping}. The diagonal line corresponds to perfect agreement.}
    \label{fig:dg-diffnet-fromsbfes}
\end{figure}

\subsection{Protein-Ligand Systems}

The relative binding free energy estimates (RBFEs) of 1-amidinopipepridine vs.\  benzamidine for the trypsin and thrombin receptors are listed in Table \ref{tab:dg-relative-pl}. The receptor swapping free energies (RSFEs) of the two compounds across the two receptors are listed in Table \ref{tab:dg-swapping-pl}. The measured $pK_i$'s of benzamidine to trypsin and thrombin are $7.51$ and $6.44$, respectively, compared to $6.44$ and $6.82$ for 1-amidinopipepridine,\cite{hilpert2002design} indicating that benzamidine binds moderately more strongly to trypsin than 1-amidinopipepridine and that the two compounds bind thrombin with nearly equal strength. The calculated RBFEs agree with the differences of experimental binding free energies derived from the inhibition constants ($\Delta G_b^\circ = -k_B T pK_i \ln 10$), although the relative preference of benzamidine for trypsin is slightly overestimated by approximately $0.7$ kcal/mol (Table \ref{tab:dg-relative-pl}).

The calculated RSFEs shown in Table \ref{tab:dg-swapping-pl} are consistent with each other and are in good agreement with the differences between the RBFEs and with the experiments. To validate convergence and any potential hysteresis, the RSFEs have been calculated in both directions: from benzamidine bound to trypsin and amidinopipepridine bound to thrombin and the other way around. The deviation between the two estimates is $0.27$ kcal/mol, which is within statistical uncertainty, and confirms the reliability of the receptor swapping alchemical protocol. Eq.\ (\ref{eq:swap-rbfe}) and the RBFEs in Table \ref{tab:dg-relative-pl} give a value of $2.52$ kcal/mol for the free energy of swapping the receptors starting from trypsin bound to benzamidine and thrombin bound to 1-amidinopipepridine, a value that is within statistical uncertainty of the result of $2.05$ kcal/mol with direct swapping alchemical protocol and $-2.32$ kcal/mol for the reverse process. The agreement observed is encouraging because the RBFE and RSFE protocols are based on distinct alchemical processes. For example, RBFE calculations involve the simulation of the solvated state of each ligand, whereas these states are bypassed in the receptor swapping protocol. 

Similarly to the host-guest results, Table \ref{tab:dg-swapping-pl} shows that the standard deviation of the RSFE from the direct swapping process is significantly smaller than that from the differences of RBFEs ($0.28$ kcal/mol and $0.45$ kcal/mol, respectively). Assuming independent Gaussian-distributed fluctuations, it would take two and a half times longer RBFE simulations to reach the same level of convergence as the direct process. Considering the need for two RBFE calculations, the direct receptor swapping process is approximately five times more computationally efficient for estimating changes in selectivity coefficients than the RBFE protocol. Some added efficiency comes from avoiding the accumulation of statistical error when taking the difference between RBFEs. However, as evident from the larger RBFE standard deviations in Table \ref{tab:dg-relative-pl} compared to those of RSFEs in Table \ref{tab:dg-swapping-pl}, the variance of the RBFE process is intrinsically larger than that of the RSFE process probably due to the added statistical fluctuations related to the solvated ligand states.  Likely due to slow apo to holo receptor reorganization effects,\cite{chen2023enhancing} we were not able to obtain converged absolute binding and receptor hopping free energies for the trypsin/thrombin complexes within a similar simulation timescale as the RBFE and RSFE calculations, indicating that those protocols would be significantly less efficient for relativity selectivity predictions than the RBFE and RSFE protocols.  

The measured affinities\cite{hilpert2002design} translate into a selectivity coefficient of $9.7$ for benzamidine in favor of binding to trypsin over thrombin. In contrast, 1-amidinopipepridine is slightly more selective (a $2.4$ selectivity coefficient) for thrombin over trypsin. Hence, benzamidine is experimentally determined to be $9.7/(1/2.4) \simeq 24$ times more selective for trypsin over thrombin than 1-amidinopipepridine. The corresponding selectivity coefficient ratio calculated from the average of the calculated RSFEs in Table \ref{tab:dg-relative-pl} and Eq.\ (\ref{eq:sel-coeff-ratio-from-RSFE}) is $81$, which is in reasonably good agreement with the experiments.

\begin{table}
\caption{\label{tab:dg-relative-pl}   ATM estimates of the relative binding free energies (RBFEs) for 1-amidinopiperidine (Am) and benzamidine (Bz) to trypsin and thrombin compared to experimental relative affinities computed from individual binding affinities.}
\begin{center}
\sisetup{separate-uncertainty}
\begin{tabular}{l S[table-format = 3.2(2)] S[table-format = 3.2(2)]}
  & \multicolumn{1}{c}{$\Delta G_r$$^{a,b}$} & \multicolumn{1}{c}{$\Delta G_r$$^{a,c}$} \\ [0.5ex]
  & \multicolumn{1}{c}{(RBFE)} & \multicolumn{1}{c}{(experimental)} \\
     \hline
Thrombin, Am to Bz   &  0.33(32) & 0.41 \\ 
Trypsin, Bz to Am    & 2.19(32) & 1.46 \\  
\hline
\end{tabular}
\begin{flushleft}\small
$^a$In kcal/mol. $^b$Uncertainties are reported as twice the standard deviation. $^c$From reference \citenum{hilpert2002design}.
\end{flushleft}
\end{center}
\end{table}

\begin{table}
\caption{\label{tab:dg-swapping-pl}  ATM receptor hopping free energy estimates (RSFEs) for 1-amidinopiperidine (Am) and benzamidine (Bz) to trypsin and thrombin compared to the corresponding differences of RBFEs from Table \ref{tab:dg-relative-pl}  and the experimental relative selectivities computed from individual binding affinities.}
\begin{center}
\sisetup{separate-uncertainty}
\begin{tabular}{l S[table-format = 3.2(2)] S[table-format = 3.2(2)] S[table-format = 3.2(2)]}
  & \multicolumn{1}{c}{$\Delta G_s$$^{a,b}$}   & \multicolumn{1}{c}{$\Delta G_s$$^{a,b,c}$} &
        \multicolumn{1}{c}{$\Delta G_s$$^{a,c}$} \\ [0.5ex]
      & \multicolumn{1}{c}{(RSFE)} & \multicolumn{1}{c}{(from RBFEs)} &
        \multicolumn{1}{c}{(experimental)} \\ [0.5ex]
\hline
\parbox{2 in}{\raggedright \singlespacing Swapping trypsin with Bz with thrombin with Am}   & 2.37(30) & 2.52(45) & 1.87 \\ 
\parbox{2 in}{\raggedright \singlespacing Swapping of trypsin with Am with thrombin with Bz}  & -2.88(30) & -2.52(45)  & -1.87 \\ 
\multicolumn{4}{c}{}\\
\hline
\end{tabular}
\begin{flushleft}\small
$^a$In kcal/mol. $^b$Uncertainties are reported as twice the standard deviation. $^c$From Table \ref{tab:dg-relative-pl}.
\end{flushleft}
\end{center}
\end{table}

\section{Discussion}

The selectivity profile of a drug candidate can be as critical as the raw strength of binding to the desired target. Often, to avoid side effects, it is desirable to inhibit a particular member of a protein family\cite{lu2012overview,bayly1999structure} or to tune the activity of a set of receptors differently than those of related isoforms.\cite{gadhiya2018new} To minimize toxicity, anti-viral and antibiotic compounds are designed to target viral and bacterial proteins while sparing the host's receptors and enzymes.\cite{li2020amide,dalhoff2021selective} Drugs used in selective cancer therapies, especially, are designed to target specific protein mutations without significantly disrupting the function of the wild-type forms.\cite{janes2018targeting,schonherr2024discovery} Conversely, identifying compounds with a wide but controlled activity profile is sometimes desirable to, for example, protect against the insurgence of resistance mutations.\cite{tmc125tmc278}

Binding selectivity coefficients are often measured to probe quantitatively the propensity of a compound to target one receptor over another.\cite{di2007understanding} While in medicinal work they are often measured in terms of inhibition concentrations (IC50's),\cite{bayly1999structure} selectivity coefficients are formally defined as the ratio of the binding constant of a ligand to a target receptor over that to a reference receptor.\cite{klotz1997ligand} A large selectivity coefficient reflects a strong preference for the ligand for the target receptor. The difference between the standard binding free energies of the ligand to the two receptors (termed here the binding selectivity free energy, BSFE) holds equivalent information to the selectivity coefficient.\cite{albanese2020structure} For example, a large and negative BSFE value reflects a strong preference of the ligand toward the target receptor. 

Computationally, binding free energies are often studied using alchemical molecular simulations. While challenging for compounds the size of common drugs, there are existing ABFE alchemical computational protocols that can yield the standard binding free energies of molecular complexes.\cite{Mobley2006,GallicchioSAMPL4,chen2023enhancing} Hence, a computational route for evaluating a BSFE consists of computing the ABFE of a ligand to two receptors separately and taking the difference in the results. Similarly, some alchemical free energy implementations support the calculation of free energy changes resulting from the mutation of one protein residue into another. When combined into a thermodynamic cycle involving bound and free forms of the receptor, these calculations probe the effect of single-point mutations on the binding free energies of protein-ligand complexes.\cite{panel2018accurate,clark2019relative,aldeghi2019accurate} 

Computational methods that yield selectivity coefficients directly can offer advantages over methods that obtain such metrics from the combination of multiple simulation results or are limited to single-point mutations. This work shows that BSFEs can be calculated directly using a receptor hopping protocol whereby a ligand is transferred from one receptor to another in a single simulation. One obvious advantage is simplifying the computational workflow with fewer calculations to set up, conduct, and analyze. A more significant advantage that we intend to explore in future work is the possibility that receptor hopping calculations for protein-ligand systems converge faster than the difference of the ABFEs. We expect this to be the case primarily because receptor hopping calculations bypass the solvated state of the ligand, where it could reorganize into conformations incompatible with binding.\cite{khuttan2024make} In contrast, both ABFE calculations involved in the estimation of a BSFE by difference would have to converge the solvated state of the ligand that, since it is the same for both receptors, is irrelevant to binding selectivity analysis. The small and rigid molecular ligands studied here do not fully probe this aspect of the method because they do not extensively reorganize upon binding. In future work, we intend to study more complex ligands\cite{khuttan2024make} and explore the effect of ligand reorganization on the efficiency of receptor hopping calculations of selectivity coefficients relative to alternative strategies.

In addition to bypassing the solvated state of the ligands, the receptor hopping strategy applies to receptor pairs differing by more than single-point mutations. For example, in the present application, we successfully modeled the ligand selectivity against the TEMOA and TEETOA hosts that differ in the methylation of multiple side-chains. We expect that the receptor hopping strategy applies to arbitrary receptor pairs as long as the dimensions and structure of their binding sites are approximately the same.

While it addresses some shortcomings of ABFE protocols, the receptor hopping protocol does not fully resolve their limitation to small ligands. Receptor hopping simulations for large ligands are expected to suffer the same difficulties encountered with ABFE and hydration free energy estimation.\cite{khuttan2021alchemical} The fundamental reason is that introducing a large molecule into a receptor binding site, whether transferred there from a vacuum, as in hydration free energy calculations, or from solution, as in ABFE calculations, or from another receptor, as in receptor hopping calculations, constitutes a severe perturbation of the system that is difficult to model. For example, the receiving receptor would have to reorganize, and any water molecules present within the receptor binding pocket would have to move into the solvent bulk to make space for the ligand.\cite{azimi2022application}  

To address this limitation, in this work, we show that estimating the BSFEs of a set of ligands for two receptors is feasible by a generalized DiffNet\cite{xu2019optimal} analysis of the results of receptor swapping free energy calculations (RSFEs). In a receptor swapping simulation, one ligand is transferred from one receptor to another while the other is simultaneously transferred in the opposite direction. RSFE calculations are expected to converge rapidly because, in addition to bypassing the solvated states of the ligands, as in receptor hopping calculations, they also bypass the apo solvated states of the receptors. Each receptor sees one ligand replaced by the other without experiencing the apo state. If the two ligands are sufficiently similar, the perturbation caused by a receptor swapping process is expected to be significantly less severe than those of ABFE and receptor hopping processes. 

In this work, we have benchmarked the receptor swapping route to BSFEs and confirmed its correctness in the case of small guests amenable to ABFE and receptor hopping calculations. We also successfully validated the RSFE protocol for protein-ligand complexes against RBFE calculations and experiments. In future work, we intend to further probe these ideas by tackling the calculation of BSFEs of large ligand libraries against protein receptors through DiffNet analysis of RSFE data.

The Alchemical Transfer Method (ATM)\cite{wu2021alchemical,azimi2022relative} proved an ideal computational platform for implementing the receptor hopping and swapping protocols presented here. ATM connects the unbound and bound states of the complex by a coordinate transformation that transfers the ligand from the solvent to the receptor binding site. The receptor hopping strategy is essentially the same, except that the ligand is transferred from one binding site to another. Similarly, the receptor swapping process is implemented similarly to the ATM RBFE protocol by moving one ligand from the first receptor to the second while simultaneously moving the other in the opposite direction. This work confirms the versatility of ATM's design and adds two more protocols (receptor hopping and receptor swapping) to the already established absolute binding free energy (ABFE) and relative binding free energy (RBFE) ATM protocols. 

\section{Conclusions}

We presented the receptor hopping and swapping free energy estimation protocols built upon the Alchemical Transfer Method (ATM) to study the binding selectivity of ligands across two receptors. The receptor hopping protocol estimates binding selectivity free energies directly without the need to simulate the solvated state of the ligand. The receptor swapping protocol measures the difference of binding selectivity free energies between a pair of ligands for two receptors directly without simulating the ligands' solvated states and the receptors' apo states. A generalization of the DiffNet analysis procedure combines the receptor swapping free energies for a set of ligands to estimate their binding selectivity free energies for two receptors. The novel methods introduced here and their implementations are benchmarked on simple but non-trivial host-guest and protein-ligand complexes and found to yield values consistent with experimental data and differences of absolute and relative binding free energies. Future work will test the applicability of the methods to the calculation of selectivity coefficients of larger sets of protein-ligand systems and probe their advantages and disadvantages over conventional strategies.

\section{Acknowledgements}

We acknowledge support from the National Institute of General Medical Sciences (NIH 1R15GM151708).

\section{Data and Software Availability}

The AToM-OpenMM software used in this study is publicly available at {\tt https://github.com/\-Gallicchio-Lab/\-AToM-OpenMM}. GitHub repository \url{https://github.com/Gallicchio-Lab/receptor-hopping} contains the AToM-OpenMM input files for the ABFE, RBFE, RHFE, and RSFE calculations reported in this work.


\providecommand{\latin}[1]{#1}
\makeatletter
\providecommand{\doi}
  {\begingroup\let\do\@makeother\dospecials
  \catcode`\{=1 \catcode`\}=2 \doi@aux}
\providecommand{\doi@aux}[1]{\endgroup\texttt{#1}}
\makeatother
\providecommand*\mcitethebibliography{\thebibliography}
\csname @ifundefined\endcsname{endmcitethebibliography}
  {\let\endmcitethebibliography\endthebibliography}{}

\end{document}